\documentclass{acm_proc_article-sp}

\usepackage{url}
\usepackage{hyperref}
\usepackage{listings}
\usepackage{color}
\definecolor{light-gray}{gray}{0.95}
\definecolor{mygreen}{rgb}{0,0.6,0}
\definecolor{myamber}{rgb}{1.0, 0.49, 0.0}

\begin{document}

\title{
Object-Oriented Parallel Programming
}
%

\numberofauthors{3} 
%
\author{
%
%
\alignauthor Edward Givelberg \\
       \affaddr{Department of Physics and Astronomy}\\
       \affaddr{Johns Hopkins University}\\
       \email{givelberg@jhu.edu}
}
\date{15 April 2014}

\maketitle
\begin{abstract}
We introduce an object-oriented framework for
parallel programming, which is based on the observation that
programming objects can be naturally interpreted as processes.
A parallel program consists of a collection of 
persistent processes that communicate by executing remote methods.
We discuss code parallelization and process persistence,
and explain the main ideas in the context of computations
with very large data objects.
\end{abstract}




\newpage
\section{Introduction}
\label{sec:Introduction}

Since the early 1990s OpenMP
\cite{openmp}
and MPI
\cite{mpi}
have been developed as successful frameworks for
shared memory and distributed memory programming.
Parallel programming has been a subject of extensive research
(\cite{herlihy2012art, mattson2004patterns}),
which led also to development
of many specialized high-level languages,
such as the PGAS programming languages
\cite{yelick2007productivity}.

In this paper we introduce an object-oriented framework for
parallel programming, which is based on the observation that
programming objects can be naturally interpreted as processes.
Our research has been motivated by the problem of 
performing computations with very large data sets.
We show, through a sequence of simple examples, that
the obejct-oriented framework is sufficiently flexible
for performing both local ``close to the data'' computations
and global data-intensive tasks requiring moving data
with maximal parallelism.

The increasing gap between the speed of performing arithmetic
operations and the speed of moving data has recently led
to development of communication-avoiding algorithms
\cite{demmel2008avoiding}.
On the other hand,
the problem of computing a Fourier transform
on a very large (Petascale) three-dimensional array
can be considered as a prototype problem where massive and highly
parallel data communications are necessary.
This problem is motivating the examples of this paper.
Since we currently do not have a compiler implementing 
processes,
as they are described in
\ref{sec:Processes},
our computation of the Fourier transform imitates this
framework using standard C++ and several functions of
the MPI 2.0 standard.
The results of the large-scale Fourier transform
computation will be reported in a separate paper.

We introduce the main ideas informally,
demonstrating simple use cases.
Our examples are written in C++ pseudo-code, but the ideas naturally
apply to any object-oriented language.
%
Processes are described in sections
\ref{sec:Processes}
and
\ref{sec:ProcessInheritance}.
The important topics of parallelization and process persistence are
dealt with very briefly in sections
\ref{sec:ParallelComputation}
and
\ref{sec:PersistentProcesses}, respectively.

\section{
Processes
}
\label{sec:Processes}
Programming objects can be thought of as processes.
We introduce this idea using a simple example.
Consider a device for storing blocks of unstructured data
which are described by the following {\tt Page} class:
\newpage
\begin{lstlisting}
class Page
{
public:
	Page(int n, unsigned char * data);
	~Page();
private:
	unsigned char * data;
};
\end{lstlisting}
A {\tt Page} object stores {\tt n} bytes of data in {\tt data}.
The block storage device, defined in the {\tt PageDevice} class,
uses a file to store multiple
data pages of the same size:
%
\begin{lstlisting}
class PageDevice
{
public:
	PageDevice(
		string filename,
		int NumberOfPages,
		int PageSize
	);  
	~PageDevice();
	void write(Page * p, int PageIndex);
	void read(Page * p, int PageIndex);
protected:
	string filename;
	int NumberOfPages;
	int PageSize;
private:
	FILE * f;
};
\end{lstlisting}
The implementation of this class creates a file {\tt filename} of
{\tt NumberOfPages * PageSize} bytes.
Pages of data are stored in the {\tt PageDevice} object
using a {\tt PageIndex} address, where {\tt PageIndex}
is between {\tt 0} and {\tt NumberOfPages}.
The {\tt write} method copies a data page of size {\tt PageSize}
to the location with an offset {\tt PageIndex * PageSize}
from the beginning of the file {\tt filename}.
Similarly, the {\tt read} method 
reads a page of data stored at a given integer address 
in the {\tt PageDevice}.
A new {\tt PageDevice} object is created 
as usual:
\begin{lstlisting}
int NumberOfPages = 10;
int PageSize = 1024;	// bytes

PageDevice * PageStore
	= new PageDevice("pagefile", NumberOfPages, PageSize);
\end{lstlisting}
%
Consider now the situation where multiple computers
{\em \color{mygreen}machine 0, machine 1, machine 2}, etc.
are available 
and
suppose
that the following code is executed on {\em \color{mygreen}machine 0}.
\begin{lstlisting}[escapechar=@]
PageDevice * PageStore
	= new(@\color{mygreen} \em{machine 1}@)
		PageDevice("pagefile", NumberOfPages, PageSize);

Page * page = GenerateDataPage();

int PageAddress = 17;
PageStore->write(page, PageAddress);
\end{lstlisting}
This program creates a {\tt PageDevice} object
on the remote computer {\em \color{mygreen}machine 1},
generates a page of data 
and stores it in the {\tt PageDevice} object
on {\em \color{mygreen}machine 1}.

Superficially, the above program differs from the standard C++ only in the
extension of the operator {\tt new}.
The new {\tt new} allocates objects on remote machines,
using the address of the remote machine specified inside parentheses.
This particular choice of syntax 
is not important and is only used here to illustrate the new idea.
No new syntax is needed to execute methods on remote objects.

The construction of a new object on a remote machine 
creates a new process on that machine.
This new process acts as a server which listens on a communications
port, accepts commands from the parent process, acting as a client,
and sends results back to the client.
The client-server protocol is generated by the compiler from the class
desccription.
Remote pointer dereferencing triggers
a sequence of events,
that includes several client-server communications, 
data transfer and execution
of code on both the local and the remote machines.

Process semantics extend naturally to simple
objects, as shown in the following example:
\begin{lstlisting}[escapechar=@]
double * data 
	= new(@\color{mygreen} \em{\color{mygreen}machine 2}@) double[1024];
data[7] = 3.1415;
double x = data[2];
\end{lstlisting}
When this code is executed on {\em \color{mygreen}machine 0},
a new process is created on {\em \color{mygreen}machine 2}.
This process allocates 
a block of {\tt 1024} doubles 
and deploys a server that communicates with the parent client
running on {\em \color{mygreen}machine 0}.
The execution of {\tt data[7] = 3.1415;} requires communication
between the 
\linebreak
client and the server, including sending the numbers
{\tt 7} and {\tt 3.1415} from the client to the server.
Similarly, the execution of the following command leads to
an assignment of the local variable {\tt x} with a copy of
the remote double {\tt data[2]} obtained over the network using
client-server communications.
We emphasize that code execution is sequential: 
each instruction, and all
communications associated with it, is completed before the following
instruction is executed.

Access to the {\tt data} block can be provided to several computing
processes, leading to an example of a shared memory implementation:
\begin{lstlisting}[escapechar=@]
const int N = 128;
class ComputingProcess;
ComputingProcess * computer[N];

for (int i = 0;  i < N;  i ++)
	computer[i] = new(@\color{mygreen} \em{machine i}@)
		ComputingProcess(data);
\end{lstlisting}
Although the {\tt data} block is shared among the processes,
the computation is sequential.
In section \ref{sec:ParallelComputation} we show how this computation can be parallelized.

Finally, we remark that
the notion of the class destructor in C++ extends natually to process
objects:
destruction of a remote object causes termination of the
remote process and completion
of the correspoding client-server communications.


The introduction of processes, accessible by remote pointers,
creates an object-oriented framework for parallel programming.
Processes exchange information by executing methods on remote objects 
rather than by passing messages.
Development of communication protocols, assembly and parsing of messages,
and much of the associated code optimization,
is relegated to the compiler.



\section{
Process Inheritance
}
\label{sec:ProcessInheritance}

Having defined processes as programming objects,
it is now straightforward
to derive new processes using previously defined
processes.
We illustrate process inheritance by extending the example
of the previous section.
Consider a device for storing three-dimensional
array blocks of
{\tt N1 * N2 * N3} doubles.
The {\tt ArrayPage} class below is easily derived from
the previously defined {\tt Page} class to handle
blocks of structured data.
\begin{lstlisting}
class ArrayPage:
	public Page
{
public:
	ArrayPage(
		int N1, int N2, int N3,
		double * data
	);
	double sum();
private:
	int N1, N2, N3;
}
\end{lstlisting}
We added the {\tt sum} method in the
{\tt ArrayPage} class as an example of a method that uses
the array structure of the data.
The definition of the derived process {\tt ArrayPageDevice}
is straightforward and requires no new syntax.
\begin{lstlisting}
class ArrayPageDevice:
	public PageDevice
{
public:
	ArrayPageDevice(
		string filename,
		int NumberOfPages,
		int n1, int n2, int n3
	):
		N1(n1), N2(n2), N3(n3),
		PageDevice(
			filename,
			NumberOfPages,
			N1 * N2 * N3 * sizeof(double)
		)
	{}
	double sum(int PageAddress);
private:
	int N1, int N2, int N3;
};
\end{lstlisting}
Suppose that an {\tt ArrayPageDevice} object has been created
on a remote machine using
\begin{lstlisting}[escapechar=@]
int n1 = 128;
int n2 = 128;
int n3 = 128;
ArrayPageDevice * blocks
	= new(@\color{mygreen} \em{machine 3}@)
		ArrayPageDevice(
			"array_blocks",
			NumberOfPages,
			n1, n2, n3
		);
\end{lstlisting}
The sum of all elements of the fourth page can be computed by
first copying the entire page to the local machine:
\begin{lstlisting}
int PageAddress = 4;
ArrayPage * page;
blocks->read(page, PageAddress);
double result = page->sum();
\end{lstlisting}
Alternatively, the sum can be computed on the remote machine
and only the result copied to the local machine:
\begin{lstlisting}
double result = blocks->sum(PageAddress);
\end{lstlisting}
The need to choose
between ``moving the data to the computation'' and ``moving the
computation to the data''
arises often in the context of data-intensive computations.
Object-oriented processes provide a simple mechanism for the programmer
to make the choice.

\section{
Parallel computation
}
\label{sec:ParallelComputation}
The implied semantics of processes requires the execution of
a remote method to complete before continuing with the computation.
A large computation may therefore be carried out jointly by several
machines, but no computation is carried out in parallel.
Nevertheless, parallelism can easily be achieved, as shown in the following
example.

A data-intensive application is likely to maintain a large number
of devices to store portions of a data set.
Such devices may be created using the following code:
\begin{lstlisting}[escapechar=@]
ArrayPageDevice * device[N];

for (int i = 0;  i < N;  i ++)
	device[i] = new(@\color{mygreen} \em{machine i}@)
		ArrayPageDevice(
			"array_blocks",
			NumberOfPages,
			n1, n2, n3
		);
\end{lstlisting}
The program may subsequently request to obtain pages of data
for local processing, one from each storage device:
\begin{lstlisting}[escapechar=@]
ArrayPage * buffer[N];
int page_address[N];
int k[N];
@\vdots@
for (int i = 0;  i < N;  i ++)
	device[i]->read(
		buffer[k[i]], 
		page_address[i]
	);
\end{lstlisting}
A page is copied from the address {\tt page\_address[i]}
in the {\tt i}-th device to the {\tt k[i]}-th page
in the local {\tt buffer}.
The implementation of this code is as follows:
\begin{lstlisting}[escapechar=@]
for (int i = 0;  i < N;  i ++)
{
	@\em{send}@ read @\em{command}@
		@\em{to}@ device[i] @\em{server on  \color{mygreen}machine i}@
	@\em{send}@ page_address[i]
		@\em{to}@ device[i] @\em{server on  \color{mygreen}machine i}@
	@\em{receive a page}@
		@\em{from}@ device[i] @\em{server on  \color{mygreen}machine i}@
	@\em{copy the received page to}@ buffer[k[i]]
}
\end{lstlisting}
This loop can be easily parallelized by the compiler,
by splitting it into two loops, as follows:
\begin{lstlisting}[escapechar=@]
for (int i = 0;  i < N;  i ++)
{
	@\em{send}@ read @\em{command}@
		@\em{to}@ device[i] @\em{server on  \color{mygreen}machine i}@
	@\em{send}@ page_address[i]
		@\em{to}@ device[i] @\em{server on  \color{mygreen}machine i}@
}

for (int i = 0;  i < N;  i ++)
{
	@\em{receive a page from \color{mygreen}machine i}@
	@\em{copy the received page to}@ buffer[k[i]]
}
\end{lstlisting}
When each {\tt ArrayPageDevice} in the
{\tt device} array is assigned to a different hard drive,
the processes in the above example
will carry out disk I/O in parallel.


The next example shows that 
the object-oriented model of parallel
programming has rich expressive power.
Consider a collection of processes for a joint computation 
of a Fourier transform.
\begin{lstlisting}[escapechar=@]
class Array;

class FFT
{
public:
	FFT(int myid): id(myid) { @\ldots@ }
	void SetGroup(int myN, FFT * myfft)
		{ N = myN; fft = myfft; }
	void transform(int sign, Array * a);
	@\vdots@
private:
	int N;
	int id;
	FFT * fft;
};
\end{lstlisting}
The master process creates 
{\tt N} parallel processes 
and assigns each process a unique id:
\begin{lstlisting}[escapechar=@]
FFT * fft[N];
for (int id = 0;  id < N;  id ++)
	fft[id] = new(@\color{mygreen} \em{machine id}@) FFT(id);
\end{lstlisting}
It informs each process in the group that it is a part of
a group of {\tt N} concurrent processes:
\begin{lstlisting}[escapechar=@]
for (int id = 0;  id < N;  id ++)
	fft[id]->SetGroup(N, fft);
\end{lstlisting}
Subsequent inter-process 
communication can be implemented by executing methods
on remote objects,
as shown in section
\ref{sec:Processes}.
A parallel FFT computation is carried out 
as follows:
\begin{lstlisting}[escapechar=@]
Array * a = CreateDataForTransform();

int sign = -1;		// forward
for (int id = 0;  id < N;  id ++)
	fft[id]->transform(sign, a);
\end{lstlisting}
In this computation an {\tt Array} object {\tt a} is
a complex large data object consisting of multiple processes,
exchanging information with the {\tt fft} processes
during the computation.
We give a detailed example of an {\tt Array} class in the next section.

Notice that the {\tt myfft} parameter of the {\tt SetGroup} method
is a remote pointer to an array of remote processes,
so future reference
to its members will result in additional communications.
The following deep copy implementation of
{\tt SetGroup},
which  copies the entire remote array of remote pointers
to a local array of remote pointers,
is preferable:
\begin{lstlisting}[escapechar=@]
void FFT::SetGroup(int myN, FFT * myfft)
{
	N = myN;
	fft = new FFT * [N];
	for (int i = 0;  i < N;  i ++)
		fft[i] = myfft[i];	// remote copy
}
\end{lstlisting}

Encapsulation, which is an important feature of object-orient\-ed programming,
clarifies relationships between objects,
facilitating parallelization of method execution across distinct objects.
Whenever possible, programs should be automatically parallelized by the compiler,
without the use of OpenMP-style directives.
Such parallelization may expose subtle programming bugs,
but in object-oriented programs these should be corrected 
by clarifying the objects' interfaces.

In this paper we 
only consider 
a few examples, which are all trivially parallelizable.
Parallel processes are naturally synchronized 
at the end of the {\tt for} loop,
however since these processes may be accessing common objects,
an explicit compiler-supported barrier method for arrays of objects
may be useful.
For example, the processes belonging to the {\tt fft} array 
can be synchronized with
{\tt fft->barrier();}

\section{
Persistent processes
}
\label{sec:PersistentProcesses}

In this section we develop an example of the {\tt Array} class,
which was introduced above.
The {\tt Array} class provides methods for 
computation with an array object that
requires a large number of hardware devices for its storage.
A typical example would be a half-petabyte-sized array,
stored on hundreds of hard-drives that are attached to multiple
computing nodes, which are interconnected by a fast network.
For the code in our examples to be valid
in the context of large data objects, many instances of the
{\tt int} type should be replaced by {\tt size\_t}.
Nevertheless, for simplicity we continue to use only {\tt int}
below.

The {\tt Array} class implements
a three-dimensional array of {\tt double} numbers,
indexed on the domain
\\
{\tt [0\ldots N1 - 1] * [0\ldots N2 - 1] * [0\ldots N3 - 1]}.
\\
Our storage method is to 
break up the domain into rectangular blocks of size
{\tt n1 * n2 * n3},
using an {\tt ArrayPage} object for each block.
\begin{lstlisting}[escapechar=@]
typedef
	vector<ArrayPageDevice *> 
	BlockStorage;
\end{lstlisting}
A {\tt BlockStorage} object represents the available hardware storage,
where
array data pages are stored.
A {\tt PageMap} 
maps logical array page addresses to physical addresses within a
{\tt BlockStorage} object:
\newpage
\begin{lstlisting}[escapechar=@]
typedef
	struct { int device_id; int index; }
	PageAddress;

struct PageMap
{
	virtual PageAddress
		PhysicalPageAddress(
			int i1, int i2, int i3
		) const = 0;
};
\end{lstlisting}
The {\tt device\_id} of a physical page address identifies
the
{\tt ArrayPageDevice}
in the {\tt BlockStorage} object
and the {\tt index} variable determines the page address
within the {\tt ArrayPageDevice}.
Each {\tt ArrayPageDevice} process of the {\tt BlockStorage} object
should be assigned to a different hard disk.
The {\tt PageMap} describes the array data layout and is crucial in
determining the I/O patterns of the computation.
The following class describing array subdomains will be useful
in the definition of the {\tt Array} class.
\begin{lstlisting}[escapechar=@]
class Domain
{
public:
	Domain(
		int N11, int N12,
		int N21, int N22,
		int N31, int N32
	);
@\vdots@
};
\end{lstlisting}
In the definition of the {\tt Array} class below
we provide, in addition to the constructor,
a {\tt read} and a {\tt write} methods to access a portion of
the array defined by the specified {\tt domain} object.
The {\tt Array} class is a client process for performing
computations on a small subdomain of the array data.
An application may deploy multiple coordinating {\tt Array} client
processes in parallel.
\begin{lstlisting}[escapechar=@]
class Array
{
public:
	Array(
		int N1, int N2, int N3,	
		int n1, int n2, int n3,	
		BlockStorage data,
		PageMap map
	);

	void read(
		double * subarray,
		Domain * domain
	);
	void write(
		double * subarray,
		Domain * domain
	);

	double sum(Domain * domain);

private:
	int N1, N2, N3;		// array sizes
	int n1, n2, n3;		// page sizes
	BlockStorage data;
	PageMap map;
};
\end{lstlisting}
At runtime the {\tt read} method assembles the data in {\tt subarray}
using multiple reads of {\tt ArrayPage} objects from {\tt data}.
The {\tt subarray} array should be small enough to fit within the
memory of the processor.
Similarly, the {\tt write} method updates the corresponding {\tt ArrayPage}
objects in {\tt data}.
The choice of the {\tt PageMap} determines the degree of parallelism
of these I/O operations.


The {\tt sum} method is an example of an array computation.
Its implementation uses the {\tt ArrayPageDevice::sum} method for
every {\tt ArrayPage} object corresponding to the {\tt domain}.
The partial sums are computed by the {\tt data} server processes
and combined together by the {\tt Array} client. 
The sum of the elements of the entire array can be computed by
using the {\tt Array} client in a loop over array subdomains,
and by deploying multiple {\tt Array} clients in parallel.
As mentioned before, the {\tt PageMap} determines 
the degree of parallelism of the computation,
amd its construction
should take into account that
in a large scale array computation multiple {\tt Array} processes
will run in parallel, communicating with the collection of processes
of the {\tt data} object.

The {\tt Array} example demonstrates 
a method for constructing large data sets 
and performing computations with them.
To complete the picture,
applications must be able to access previously constructed data sets.
In our view large data objects are described as collections of 
{\em persistent processes}.
Persistent processes are objects that can be destroyed only by
explicitly calling the destructor.
The runtime system is responsible for storing process representation,
and activating and de-activating processes, as needed.
Processes can be accessed
using a symbolic object address,
similar to addresses used by the Data Access Protocol (DAP)
\cite{gallagher2004data},
for example:
\begin{lstlisting}[escapechar=@]
PageDevice * page_device =
	"http://data/set/PageDevice/34";
\end{lstlisting}

Additional research is needed for implementation of process
persistence.
Here we note that
the combination of inheritance and persistence leads to interesting
use cases, such as the following example:
We add the constructor 
\begin{lstlisting}[escapechar=@]
ArrayPageDevice(PageDevice * page_device);
\end{lstlisting}
to the {\tt ArrayPageDevice} class,
so that
a new process with a pointer to an existing process can be created
as follows:
\begin{lstlisting}[escapechar=@]
ArrayPageDevice * new_device = 
	new ArrayPageDevice(page_device);
\end{lstlisting}
The {\tt new\_device} process may co-exist and communicate with the
{\tt page\_device} process,
or it may use a copy constructor to copy the state of {\tt page\_device}
and
subsequently shut it down using:
\begin{lstlisting}[escapechar=@]
delete page_device;
\end{lstlisting}

\section{
Conclusion
}
In this paper we have shown that
programming objects have a natural interpretation as processes,
and have described 
the resulting object-oriented framework for parallel programming.
In our view a parallel program consists of a collection of
persistent processes, which, in general, represent different
programming objects.
The processes communicate by executing methods on remote objects.
The resulting framework is rich enough to include
%
shared memory and distributed memory programming,
as well as other programming models
(client-server applications, map-reduce, etc.).

Processes can be added to any object-oriented language,
and should be useful in computations with large data sets,
operating system design and scientific applications.
In our opinion the process-oriented programming style
facilitates creating automatically parallelizable code.
In this paper we touched only briefly on the important
subjects of code parallelization and process persistence.
These topics, as well as issues of implementation 
and optimization require further research.

\section{
Acknowledgement
}
This research was partially funded by a grant from Intel Corp.


\bibliographystyle{abbrv}
\bibliography{paper}

\end{document}